\documentclass{article}
\usepackage{rotating}
\usepackage{amsmath}
\usepackage{natbib}
\usepackage{color}
\usepackage{gensymb}
\usepackage{lineno}

\usepackage{caption}
\usepackage{subcaption}
\usepackage{graphicx}

\usepackage[skip=10pt]{parskip}

\begin{document}


{\centering

\Large{\textbf{Latitudinal Variation of Clouds' Structure Responsible for Venus' Cold Collar}}\\

\vspace{1cm}

\large{I. Garate-Lopez\footnote{Corresponding author. Email address: itziar.garate-lopez@lmd.jussieu.fr} and S\'ebastien Lebonnois} \\

\vspace{0.3cm}

\small{\textit{Laboratoire de M\'et\'eorologie Dynamique (LMD/IPSL), Sorbonne Universit\'es, UPMC Univ Paris 06, ENS, PSL Research University, Ecole Polytechnique, Universit\'e Paris Saclay, CNRS, Paris, France}}

\par}				

\vspace{0.3cm}
 \par\noindent\rule{\textwidth}{0.5pt}

\begin{abstract}

Global Climate Models (GCM) are very useful tools to study theoretically the general dynamics and specific phenomena in planetary atmospheres. In the case of Venus, several GCMs succeeded in reproducing the atmosphere's superrotation and the global temperature field. However, the highly variable polar temperature and the permanent cold collar present at $60\degree - 80\degree$ latitude have not been reproduced satisfactorily yet. 

Here we improve the radiative transfer scheme of the Institut Pierre Simon Laplace Venus GCM in order to numerically simulate the polar thermal features in Venus atmosphere. The main difference with the previous model is that we now take into account the latitudinal variation of the cloud structure. Both solar heating rates and infrared cooling rates have been modified to consider the cloud top's altitude decrease toward the poles and the variation in latitude of the different particle modes' abundances. 

A new structure that closely resembles the observed cold collar appears in the average temperature field at $2\times10^{4} - 4\times10^{3}$~Pa ($\sim62 - 66$~km) altitude range and $60\degree - 90\degree$ latitude band. It is not isolated from the pole as in the observation-based maps, but the obtained temperature values (220~K) are in good agreement with observed values. Temperature polar maps across this region show an inner warm region where the polar vortex is observed, but the obtained 230~K average value is colder than the observed mean value and the simulated horizontal structure does not show the fine-scale features present within the vortex. 

The comparison with a simulation that does not take into account the latitudinal variation of the cloud structure in the infrared cooling computation, shows that the cloud structure is essential in the cold collar formation. Although our analysis focuses on the improvement of the radiative forcing and the variations it causes in the thermal structure, polar dynamics is definitely affected by this modified environment and a noteworthy upwelling motion is found in the cold collar area. 

\end{abstract}

\vspace{0.3cm}
\textit{Kew words:} Venus atmosphere, cold collar, modelling.


%
%


\newpage

\section{Introduction}\label{Introduction}

In the last two decades, Global Climate Models (GCMs) have turned out to be very useful tools to study the general atmospheric circulation on Venus and the role of different phenomena in the angular momentum budget. Several numerical simulations by different GCMs have already reproduced the main characteristic of the Venus atmosphere: its superrotation \citep[e.g.][]{yamamototakahashi03a, lee07, sugimoto14a, lebonnois16a}. At about 70~km altitude clouds circle the solid planet at $\sim$100~m/s, sixty times faster than the rotation of the planet. These simulations support the Gierasch-Rossow-Williams mechanism \citep{gierasch75, rossowwilliams79} to explain the maintenance of the superrotation, which holds the latitudinal transport of angular momentum by horizontal planetary-scale waves and by the meridional circulation responsible for the velocity field in the atmosphere. GCMs also confirmed that in the case of Venus, the thermal tides play a key role in transporting the angular momentum vertically \citep{takagimatsuda07, lebonnois10}. On the other hand, it is well known that there is a wide variety of waves in the main cloud deck \citep{belton76, rossow80, delgeniorossow90, peralta08, piccialli14} that may contribute to the transport of angular momentum and, therefore, recent GCMs studies have focused on the analysis of the wave activity present in the Venusian atmosphere \citep{sugimoto14b, lebonnois16a}.

One of the pending questions for the Venus GCMs is the simulation of the atmospheric thermal structure. Although the average temperature field has been reproduced adequately, no model has satisfactorily simulated one of its most puzzling properties: the polar temperature distribution, formed by highly variable vortices and cold air areas that surround the vortices permanently \citep{zasova07,tellmann09}.

The temperature in the deep atmosphere is almost constant from equator to around $50\degree$, and slowly decreases polewards \citep{tellmann09}. However, above $\sim$70~km altitude temperatures increase from equator to pole, in contrast to what is expected from radiative-convective equilibrium, since solar heating is higher at equator than at the poles. This is known as the \textit{warm polar mesosphere}. At 60 - 70~km altitude the temperature field in the $60\degree - 80\degree$ region has a striking feature in both hemispheres, a strong temperature inversion called the \textit{cold collar} \citep{taylor80}. It is a collar of cold air circling each pole of Venus. The latitudinally averaged temperature between $60\degree$ and $80\degree$ is about 20~K colder than at the equator and 15~K colder than at the poles, i.e. between $80\degree$ and $90\degree$ \citep{haus14}. This temperature difference would cause it to dissipate rapidly, however the cold collar is a permanent structure at the sub-polar latitudes of Venus, implying that it is forced by some unknown mechanism. Enclosed by the cold collar there is a \textit{warm polar vortex} in each pole of Venus. These are rapidly rotating cloud and temperature structures that show highly variable and complex morphologies with warm small-scale filaments \citep{piccioni07,luz11,garatelopez13}. This region features the greatest horizontal temperature gradient at the cloud top level. Neither the cold collar nor the warm vortex are completely uniform along the latitude circles \citep{grassi10, garatelopez15}; therefore, the latitudinally averaged temperature field hides the real thermal contrast between these two features. The mean temperature difference between the coldest area of the cold collar and the warmest nucleus of the vortex is $\sim$30~K, but the temperature gradient can be as high as $dT/dr$ = 0.1~K/km ($\delta T$ = 50~K over 500~km) in this region \citep{garatelopez15}.

Although these polar structures have been observed repeatedly, their origin and formation mechanism are still unknown and, therefore, numerical studies of the Venusian polar region were recently conducted \citep{yamamototakahashi15, ando16, ando17, lebonnois16a}. Using GCMs that force the temperature structure with Newtonian cooling and previously prepared heating rate profiles, \cite{yamamototakahashi15} and \cite{ando16} concluded that thermal tides play an important dynamical role in the formation of the cold collar and warm vortex. The former stated that the baroclinic waves are also part of the formation mechanism while the latter attributed the origin of the polar structures to the combination of the thermal tides and the residual meridional circulation. \cite{ando17} investigated the vertical structure of the temperature fluctuation in the Venusian polar atmosphere by comparing radio occultation measurements and GCM results, and concluded that the thermal disturbance associated to the polar vortex could be a neutral barotropic Rossby wave related to barotropic instability. These three works obtained a temporally evolving polar region and some horizontal structure within the vortex. However, the average latitude-altitude distributions of temperature do not exactly reproduce the observed cold collar structure. In observation-based maps \citep{zasova07,tellmann09,haus14} the cold collar is an isolated region and is located slightly lower than in the maps of \cite{yamamototakahashi15} and \cite{ando16, ando17}. In addition, the average temperature values obtained by \cite{ando16} are not warm enough compared to observations.

On the other hand, \cite{lebonnois16a} used the IPSL (Institut Pierre Simon Laplace) Venus GCM, that has a full radiative transfer module, to model the Venus atmosphere paying a special attention to polar regions. Their zonally and temporally averaged latitudinal profiles showed a small cold collar signature at the cloud top ($\sim$70~km), located slightly higher than observed. Due to its axi-asymmetric distribution, the obtained cold collar looked subtle when zonally averaged but it was more evident in the polar temperature fields. They related the shape of the polar temperature distribution (cold collar and warm vortex) to the combination of thermal tide and high-frequency wave activity but the temperature contrasts obtained inside the polar regions ($\sim$10~K) were weaker than those presented by \cite{ando16, ando17} ($\sim$20-25~K) and were far from the observed $\sim$30~K on average \citep{garatelopez15}.

In order to perform an in-depth study of the Venus polar atmosphere and try to better understand the origin and nature of the cold collar and warm vortex, we have adapted the IPSL Venus GCM by taking into account the latitudinal variation of the cloud structure (not considered in any GCM mentioned above). 
The details about the improvements on the radiative transfer scheme are described in Section \ref{Simulations}. Then, Section \ref{Results} shows the average temperature and zonal wind fields obtained by the current model, which are in better agreement with observations and reproduce the characteristic cold collar feature in a more realistic way than previously. We discuss the formation of the cold collar in terms of radiative transfer effects in Section \ref{ColdCollarAnalysis} and we sum up the conclusions in Section \ref{Conclusions}.


\section{Simulations}\label{Simulations}


\subsection{IPSL Venus GCM}\label{IPSLVenusGCM}

The general characteristics of the IPSL Venus GCM have been described in detail in \cite{lebonnois10} and \cite{lebonnois16a}. Therefore, only a short summary is given here. The model is based on the LMDZ latitude-longitude grid finite-difference dynamical core \citep{hourdin06}. Since the goal of the current study is the polar regions, it must be noted that this dynamical core includes a longitudinal polar filter that limits the effective resolution to that at mid-latitudes. Some of the most important features of the model are: the boundary layer scheme used, based on \citet{melloryamada82}; the implementation of topography, with hybrid vertical coordinates (50 vertical levels, from surface to roughly 95~km altitude); the temperature dependence of the specific heat $C_p(T)$, which affects the definition of the potential temperature \citep[as detailed in][]{lebonnois10}; and the horizontal resolution currently used, 96 longitudes x 96 latitudes, i.e. 3.75\degree~x~1.875\degree.

The previous version of the model \citep{lebonnois16a} reproduced consistently the average zonal wind and temperature fields, but the obtained equatorial jet was too intense compared to mid-latitude jets and the winds below the clouds were too slow compared to measurements made by Venera and Pioneer Venus probes \citep{schubert83,gierasch97}. On the other hand, the modeled average temperature structure, although generally consistent with observations, resulted in colder temperatures in the deep atmosphere and at surface compared to observed profiles \citep{seiff85}, and in slightly higher temperatures above the clouds \citep{tellmann09, migliorini12, grassi14}. The latitudinal profiles showed a small cold collar signature at $\sim$70~km and the polar temperature distribution at $\sim$67~km displayed some inner structure. However, in the average altitude-latitude temperature map the cold collar was not an isolated feature as in observation-based maps \citep{tellmann09, haus14} and the contrast between cold collar and warm vortex was weaker than observed, about 10~K in the model vs $\sim$30~K in temperature retrievals from VIRTIS instrument \citep{garatelopez15}. The latitudinal structure of the clouds was not taken into account in this previous IPSL model, though \citet{lebonnois16a} noted that it could play a role in the shape and strength of the cold collar.

Due to the development of different tools necessary to implement the latitudinal variation of the cloud structure in the solar heating rates and in the infrared cooling rates, the improvements in the radiative transfer were done in two stages; (1) implementation of the cloud structure in the solar heating rates, plus some tuning, and (2) implementation of the cloud structure in the IR cooling rates (these improvements and tuning are described in the next section).
The simulation started from an already superrotating atmosphere obtained from previous simulations \citep{lebonnois16a} and it was then run for 200 Venus days (Vd, 117 Earth days) with the partially-improved radiative transfer, the one obtained after the first stage of the improvements. Two simulations were then pursued for another 100~Vd: one with latitudinally uniform clouds in the IR cooling computation (referred to as \textit{unifcld} hereon) and another one with latitudinal variations of the cloud structure for studying the cold collar (\textit{varcld} hereon). 
The former case, \textit{unifcld} simulation, is used for control by comparing it to the \textit{varcld} simulation in the discussion of the formation of the cold collar, but this simulation and its results have no physical sense and should not be considered alone.
In the latter case (\textit{varcld} simulation) a stable state of the atmosphere was obtained after roughly 60~Vd of simulation, so it is the simulation that represents the current IPSL Venus GCM model. The different inputs used in each stage and simulations are summarized in Table \ref{paramsimus}. 

{\bf TABLE~\ref{paramsimus}}


\subsection{Improved radiative transfer}\label{Improvements}

Previously, the radiative transfer included solar heating rate profiles as a function of solar zenith angle taken from look-up tables based on \cite{crisp86}, and used the infrared net-exchange rate matrix formulation \citep{eymet09} with horizontally uniform opacity sources (gas and clouds).

For the current study, the main modification in the radiative transfer is related to the cloud model. Our computations now take into account the more recent cloud model described by \citet{haus14,haus15}, which is based on the most recent observations performed by the Venus-Express spacecraft. In particular, this model describes latitudinal variations of the cloud structure. According to \cite{haus14}, the cloud top altitude (defined as the altitude where the optical depth equals the unity at 1~$\mu$m) decreases slowly from about 71~km at the equator to about 70~km at $50\degree$, but then it quickly drops poleward reaching about 61~km over both poles. Authors also found a strong latitude dependence of cloud opacity, which is transposed into the latitudinal behavior of cloud mode factors, scaling the abundance of the different modes compared to the equatorial vertical distribution. Retrieved mode 1 and 2 factors gradually decrease poleward of $30\degree$ showing a small local minimum at $55\degree$. The mode 3 factor is nearly constant from equator to $\sim15\degree$, decreases between $15\degree$ and $30\degree$, remains almost constant up to $55\degree$, but then strongly increases poleward \citep{haus14}. This new cloud model and some tuning (see details below) were implemented both in the solar heating and infrared cooling rates' computations.


\subsubsection{Solar heating}\label{SolarHeating}

For the short-wavelength part of the radiative transfer scheme, the principle used is basically the same as previously. Look-up tables are used, that contain the vertical profile of the solar heating rate as a function of solar zenith angle. The model described in \citet{haus15} is now used to produce these tables. In addition, since the latitudinal structure of the cloud is now taken into account, the tables vary with latitude, in bins of $5\degree$. It can be noted that in this cloud model, the unknown ultraviolet absorber is modeled independently from the cloud particle modes, regardless of the absorber's chemical composition.

The vertical profile of the solar heating plays a crucial role in the temperature profile, as shown e.g. by \citet{lebonnois15}. 
However, it is poorly constrained by available data. A series of tests were performed with the new solar heating rates \citep[Methods and Supplementary Materials]{lebonnoisschubert17}. Below approximately the cloud base, these solar heating rates are smaller than the ones that were used previously \citep{crisp86}, and this resulted in deep atmospheric temperature profiles significantly colder than observed. The composition of the lower haze particles, located between the cloud base (48~km) and 30~km and observed by in-situ probe instruments \citep{knollenberg80}, is not established, so their optical properties are not well constrained. This sub-cloud haze is not taken into account in the \citet{haus15} model, except for the tail of the sulfuric acid cloud mode vertical distributions. The absorption of the solar flux in this region is therefore subject to uncertainty. To better reproduce the temperature structure in the middle cloud and below, we decided to increase the solar heating rates in this region, multiplying the solar heating rates provided by the \citet{haus15} model by a factor of 3. This tuning brings the values in the same range as other models \citep{crisp86,leerichardson11b}, and allows the temperature profile to reach observed values.


\subsubsection{Infrared cooling}\label{IRCooling}

The long-wavelength part of the radiative transfer scheme is based on the net-exchange rate (NER) formalism \citep{eymet09}.
The cloud model (density profile of each mode, $n_i(z)$) used in the previous computation of the net-exchange rate matrices was from \citet{zasova07}. 
In this work, we now use the cloud model described in \citet{haus14}.
In order to take into account the latitudinal variations of this model, matrices are computed for five latitudinal bins: $0\degree$ - $50\degree$, $50\degree$ - $60\degree$, $60\degree$ - $70\degree$, $70\degree$ - $80\degree$ and $80\degree$ - $90\degree$. When running the GCM, the matrices are interpolated between the central latitudes of each bin.

The gas opacities are computed first for high-resolution spectra, taking into account updated spectral dataset, Voigt line shape profiles, and a truncation at 200~cm$^{-1}$. Then correlated-k coefficients are computed and used in the NER matrix calculation. In addition, CO$_2$ and H$_2$O collision-induced absorption are taken into account, as detailed in \citet{lebonnois15}.
However, as mentioned in \citet{lebonnois15} and \citet{lebonnoisschubert17}, it is difficult to have consistent temperature profiles in the middle and lower cloud and in the atmosphere below without including some small additional continuum to close the windows located in the [3 - 7]~$\mu$m range, through which energy is exchanged between the cloud base and the layers just below. Here, an additional continuum of $1.3\times10^{-6}$cm$^{-1}$amagat$^{-2}$ is taken into account in the 30 - 48~km region, and $4\times10^{-7}$cm$^{-1}$amagat$^{-2}$ in the 16 - 30~km region, for a best fit of the VIRA \citep{seiff85} and VeGa-2 \citep{linkin87,zasova06} temperature profiles \citep[see][Methods and Supplementary Information]{lebonnoisschubert17}.


\section{Results}\label{Results}


\subsection{Average temperature field}\label{ResultsAvgTemp}

\begin{figure}
\centering
\includegraphics[width=9.cm]{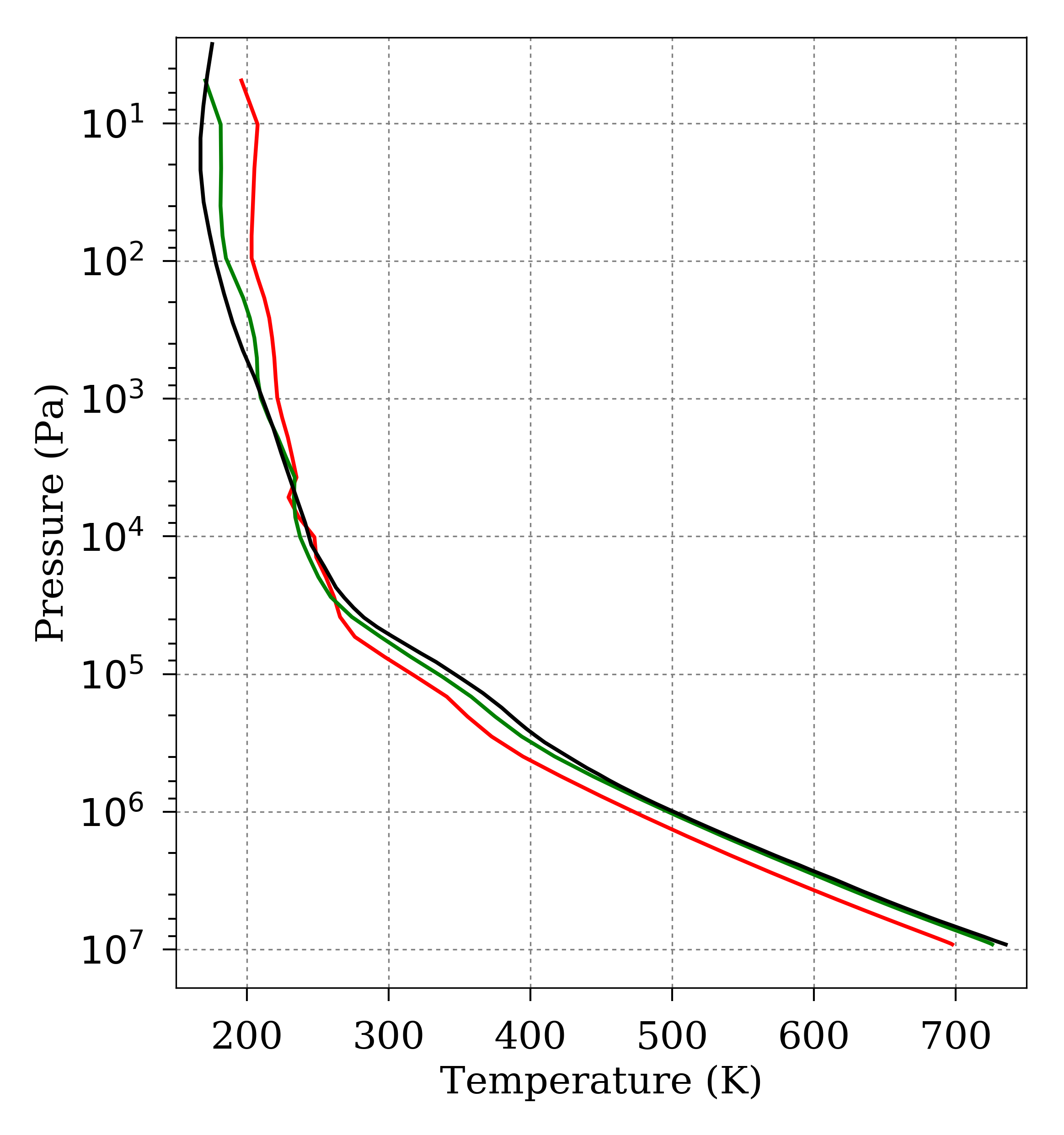}
\caption{Horizontally and temporally averaged temperature profile for the previous (red) and current (green) IPSL Venus GCM models at the end of simulation (300~Vd) compared to VIRA reference profile (black). Data is globally averaged over all longitudes and latitudes in each altitude, and over 2~Vd in time.}
\label{VertTempProf}
\end{figure}

As a direct consequence of the above described modifications done in the radiative transfer (considering both stages), the average temperature field has significantly improved. The horizontally and temporally averaged temperature profile is now (\textit{varcld} simulation mentioned above) closer to the VIRA profile (see Figure \ref{VertTempProf}) with warmer modeled temperatures below the clouds (and colder above) than in the previous model \citep{lebonnois16a}. In the 3-dimensional simulations presented here, and compared to the 1-dimensional tests performed in \citet{lebonnoisschubert17} (Methods and Supplementary Information), the average temperature field is still approximately 10~K colder than observed below roughly 70~km. However, due to the sensitivity to the solar heating rates below the clouds and the associated uncertainties, no additional tuning was performed. 

{\bf FIG~\ref{VertTempProf}}

\begin{figure}
\centering
\includegraphics[width=13.5cm]{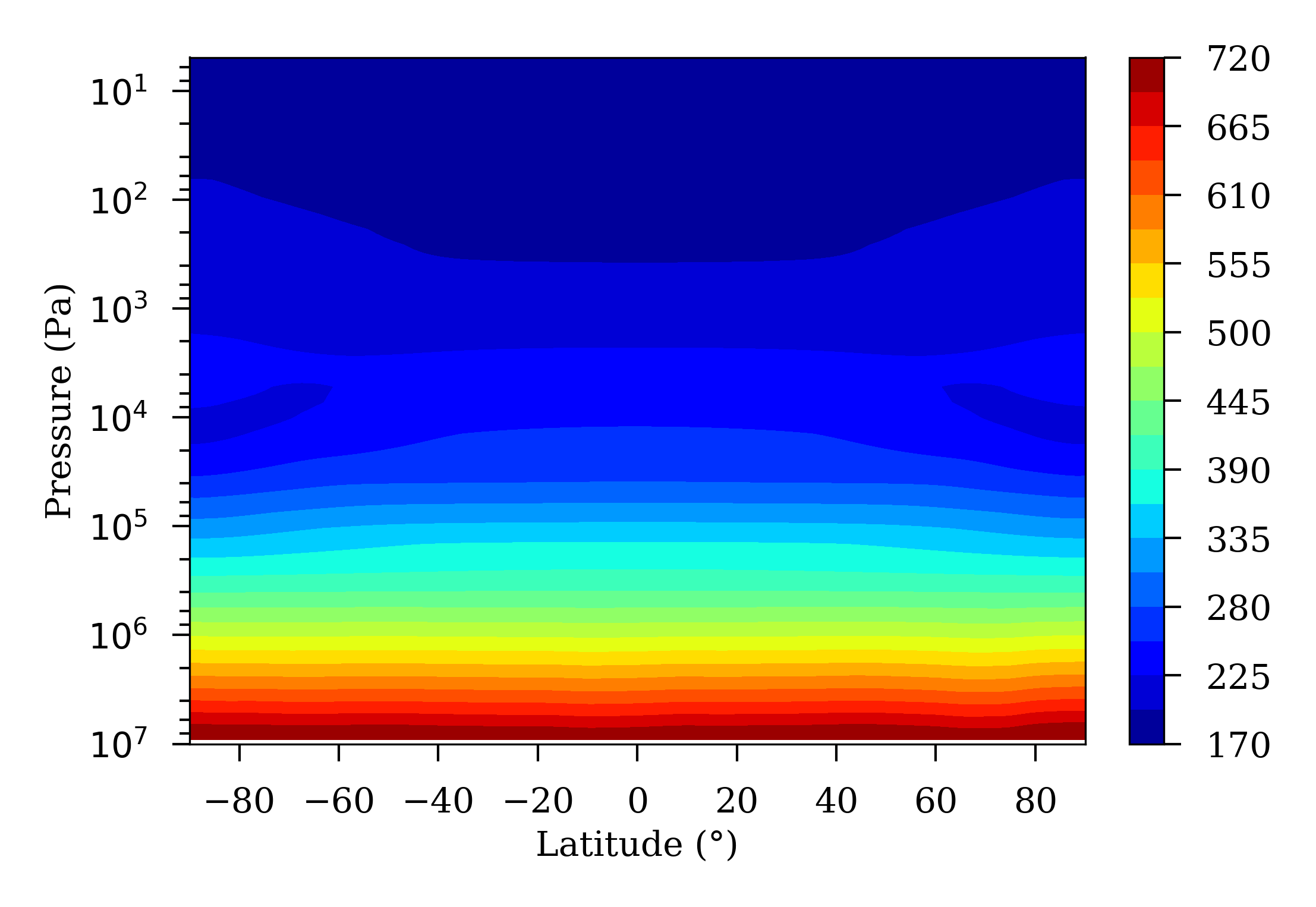}
\caption{Zonally and temporally averaged temperature field (K) for the current IPSL Venus GCM model at the end of simulation (300~Vd). Data is averaged over $360\degree$ in longitude and 2~Vd in time.}
\label{VertTempMap}
\end{figure}

The altitude-latitude distribution of the zonally and temporally averaged temperature (Figure \ref{VertTempMap}) shows two clearly differentiated behaviors. Above $4\times10^{3}$~Pa the temperature increases from equator to pole, but below $2\times10^{4}$~Pa it smoothly decreases towards the pole, just as seen in observation-based maps  \citep[e.g.][]{tellmann09,haus14}. Moreover, a new structure that resembles clearly the cold collar appears for the first time in the IPSL Venus GCM, at $2\times10^{4} - 4\times10^{3}$~Pa ($\sim62 - 66$~km) altitude and between $60\degree$ and $90\degree$ in both hemispheres. This feature is not isolated from the pole as in the observation-based maps, but it is the most similar feature reproduced so far. The temperature values obtained with our model ($\sim$220~K) are in good agreement with observations. Therefore, the latitudinal variation of the clouds' structure plays a key role in the formation of the cold collar. The development of the cold collar related to the radiative transfer in the polar atmosphere of Venus is discussed in detail below (Section \ref{ColdCollarAnalysis}).

{\bf FIG~\ref{VertTempMap}}


\subsection{Cold collar}\label{ResultsColdCollar}

\begin{figure}
\centering
\includegraphics[width=9.cm]{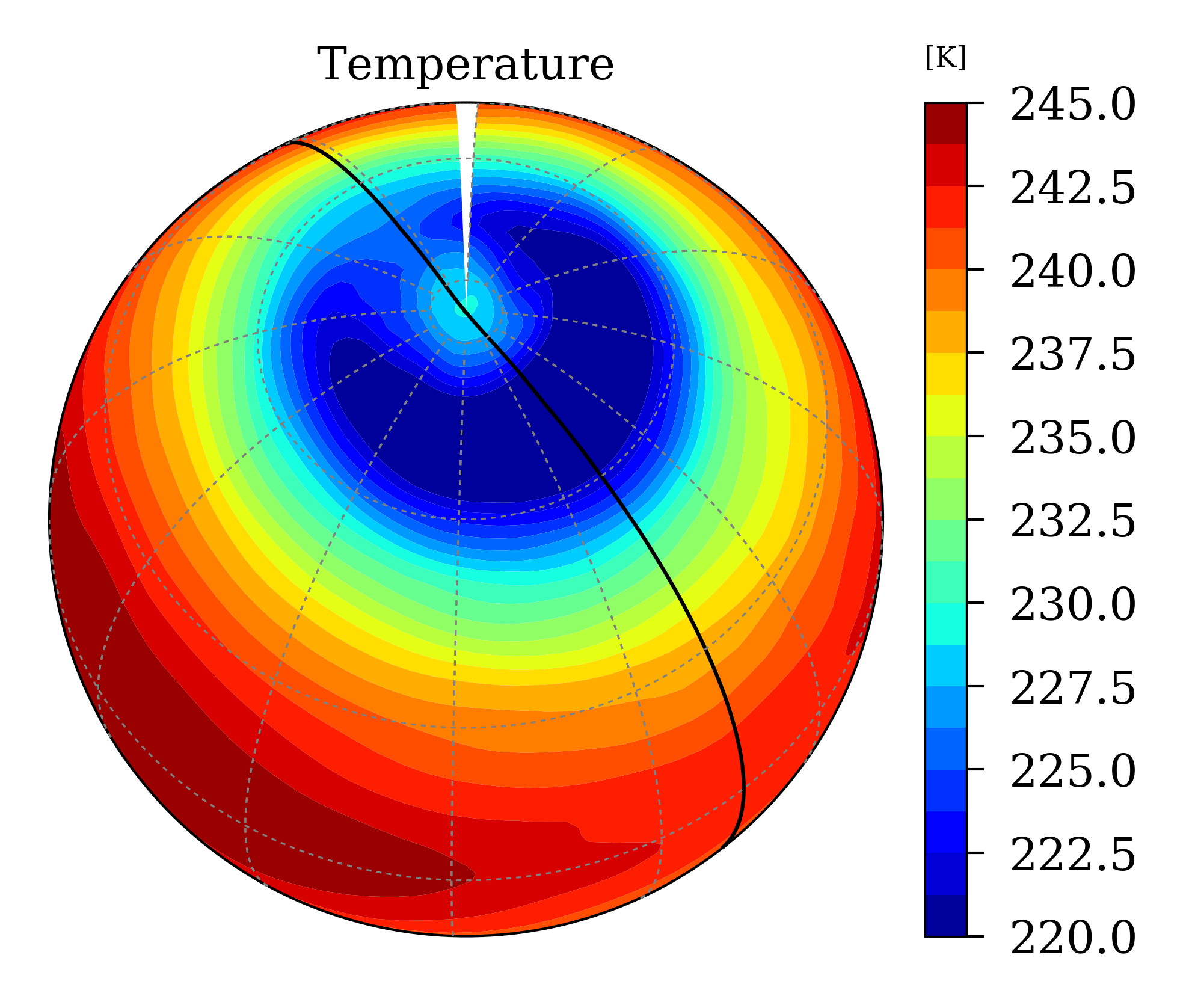}
\caption{Example of polar temperature field at $7\times10^{3}$~Pa ($\sim$64~km) for the current IPSL Venus GCM model. Temperature is averaged over 1/24~Vd $\sim$ 4.88 Earth days. Black line shows the terminator with dayside on front.}
\label{PolarTempField}
\end{figure}

Figure \ref{PolarTempField} shows an example of the temperature field at the northern pole at $7\times10^{3}$~Pa ($\sim$64~km) altitude. The atmosphere between $60\degree$ and $85\degree$ is clearly colder than the rest of the latitudes, being on average $\sim$20~K colder than the equator and about 10~K colder than the pole. The coldest region seen close to the terminator shows values about 220~K, in good agreement with values observed by VIRTIS \citep[e.g.][]{haus14, garatelopez15}. The inner warm core shows an average value of 230~K (colder than the observed mean values). However, we find values as high as 240~K in about $3\%$ of the time over the last simulated 2~Vd.

{\bf FIG~\ref{PolarTempField}}

The temporal evolution of the polar temperature obtained by our current model does not always show an inner region surrounded by the cold collar and the structure simulated within the inner core is smoother than obtained by \cite{yamamototakahashi15} and \cite{ando16, ando17} (these GCMs use spectral dynamical cores that do not have singularities at the pole). The altitude level where the cold collar is reproduced ($\sim$64~km) and the temperature values ($\sim$220~K) obtained by the IPSL Venus GCM are, however, more consistent with observations (e.g. $\sim$62~km and $\sim$220~K by \citealt{tellmann09}).

The coldest region shows a rotation around the pole with a period of about 5.85 Earth days in the model, that agrees with the 5 - 10 Earth days period found by \citep{luz11} for the drift of the rotation center of the vortex around the pole. However, the modeled inner core does not show a clear rotation around the pole, so it is difficult to compare its motion to the one observed for the warm vortex \citep{luz11, garatelopez13}.

Figure 3 in \cite{lebonnois16a} displayed four temperature fields at pressure $3\times10^{3}$~Pa ($\sim$67~km) in the northern polar region that showed a cold polar region with an inner warmer core. The altitude level where these features are obtained is slightly different comparing the previous and current models but the horizontal structure simulated in both cases is very similar in general, probably due to the longitudinal polar filter of the model that does not allow to reproduce fine-scale features.


\subsection{Average wind field}\label{ResultsAvgWind}

\begin{figure}
\centering
\includegraphics[width=13.5cm]{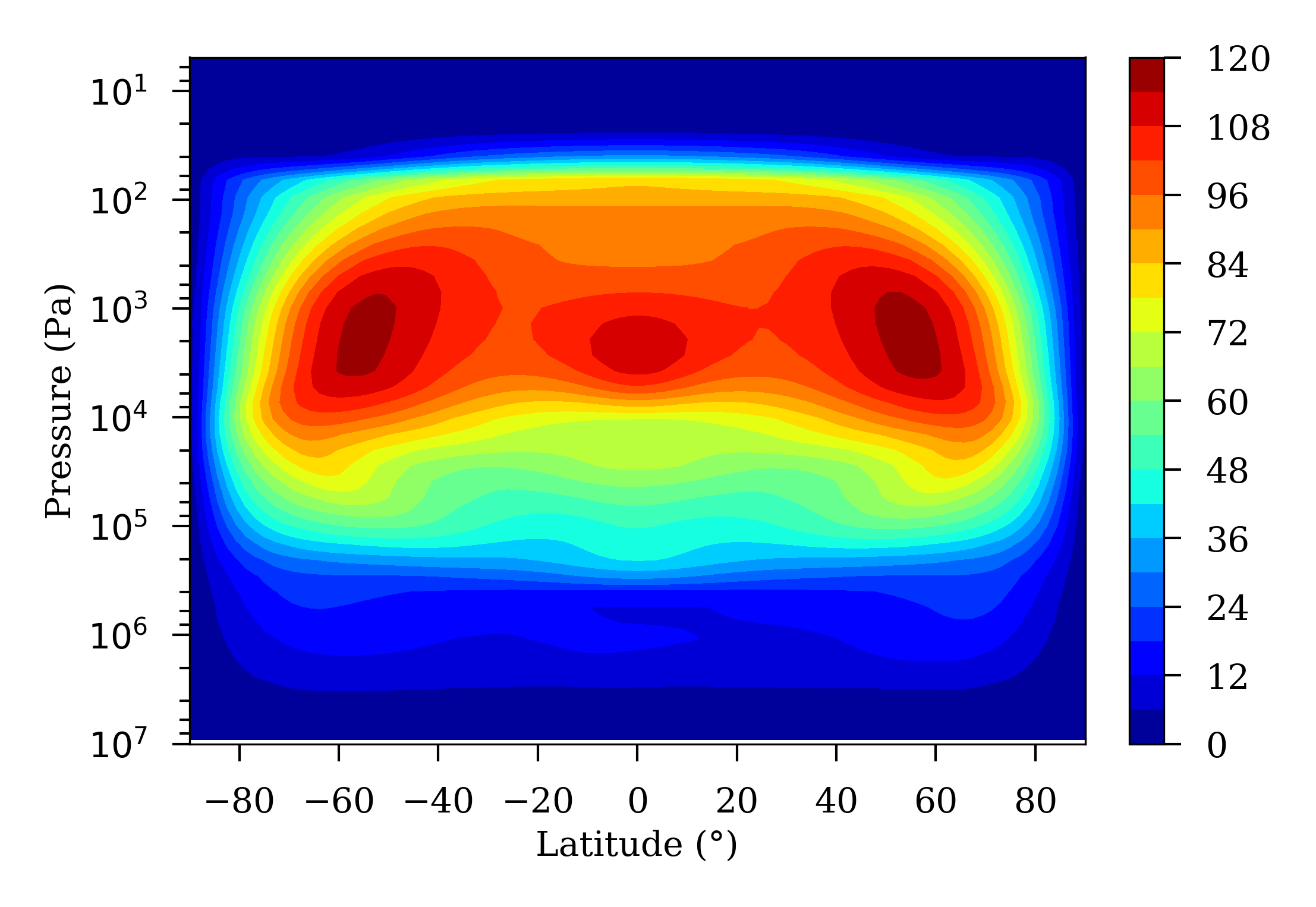}
\caption{Zonally and temporally averaged zonal wind field for the current IPSL Venus GCM model at the end of simulation (300~Vd). Data is averaged over $360\degree$ in longitude and 2~Vd in time.}
\label{VertWindMap}
\end{figure}

Considering the latitudinal structure of the clouds has also improved the zonally and temporally averaged zonal wind field. The modeled equatorial jet is now less intense than the mid-latitude jets (Figure \ref{VertWindMap}) as cloud-tracking measurements suggest \citep{hueso15}. The modeled mid-latitude jets are still located too close to the poles (and, therefore, the latitudinal wind gradient is still high poleward of $70\degree$) but their altitude has gone down slightly, while the equatorial jet has ascended a bit.

{\bf FIG~\ref{VertWindMap}}

\begin{figure}
\centering
\includegraphics[width=9.cm]{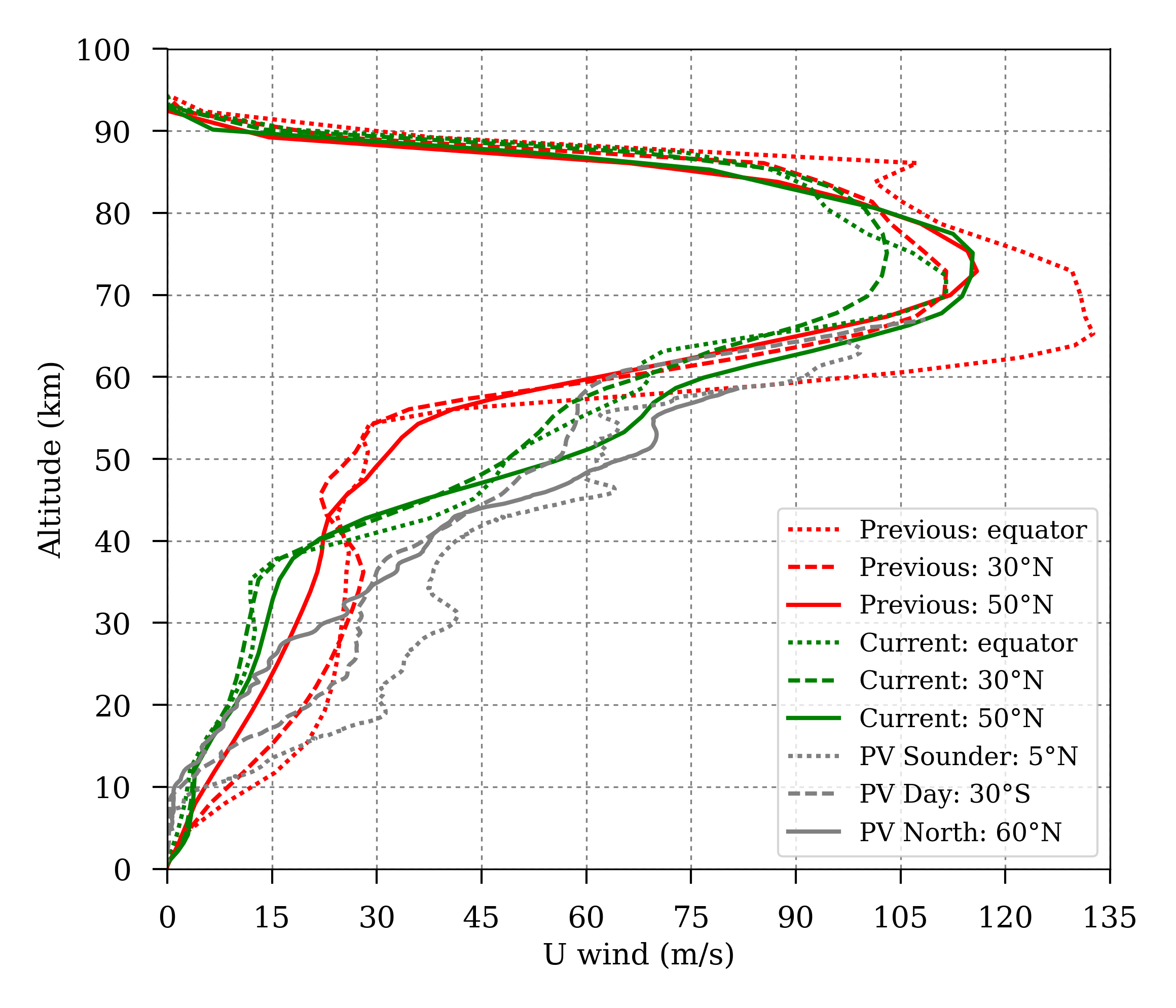}
\caption{Vertical profiles of the zonally and temporally averaged zonal wind at different latitudes for the previous (red) and current (green) IPSL Venus GCM models at the end of simulation (300~Vd). Data is averaged over $360\degree$ in longitude and 2~Vd in time. Measurements from different Pioneer Venus probes are added in gray.}
\label{VertWindProf}
\end{figure}

In the previous model \citep{lebonnois16a} winds at 40 - 60~km were approximately half of the observed values, but the currently modeled winds at this altitude range are on average $\sim20~m/s$ (about $75\%$) faster and are now in good agreement with the Venera and Pioneer Venus probes' profiles (see Figure \ref{VertWindProf}). According to a new wave activity analysis that is still on-going work and will be presented in a subsequent publication, this improvement of the averaged zonal wind could be due to a new baroclinic activity found between 40 and 60~km altitude and that transports heat and angular momentum equatorwards. Nevertheless, the velocity of the zonal winds below 40~km altitude has decreased showing winds significantly slower than observed, which was not the case previously. The large-scale gravity waves present in the deep atmosphere in the previous simulation \citep{lebonnois16a} are no longer obtained, which probably explains this worsening in the zonal wind below 40~km.

{\bf FIG~\ref{VertWindProf}}


\section{Cold Collar Analysis}\label{ColdCollarAnalysis}

To analyze in detail the influence of the latitudinal structure of the clouds in the formation of the cold collar, the simulation with the latitudinal variations of the cloud structure (\textit{varcld} simulation) is compared to a control simulation in which a uniform cloud structure is maintained in the IR cooling computation (\textit{unifcld} simulation). Note that the simulations differ in the second stage of the implementation (see Section \ref{IPSLVenusGCM}), i.e. in the last 100~Vd, so the solar heating rates take into account the latitudinal variations in both simulations (same look-up table). The \textit{unifcld} simulation has no physical sense since the cloud model is different in the solar heating and IR cooling treatments, and its results are very similar to the previous simulation presented by \cite{lebonnois16a}, but we think that the comparison between \textit{unifcld} and \textit{varcld} simulations is useful to identify which are the key elements in the formation of the cold collar.

The cold collar is formed in the \textit{varcld} simulation but not in the \textit{unifcld} simulation, which indicates that the generation of this characteristic feature is not affected by latitudinal variations of the clouds in the solar heating rates. It is the influence that the cloud structure has in the IR cooling rates that forms the cold collar.

\begin{figure}
\centering
  \begin{subfigure}[t]{0.45\textwidth}
    \centering
    \includegraphics[width=5.5cm]{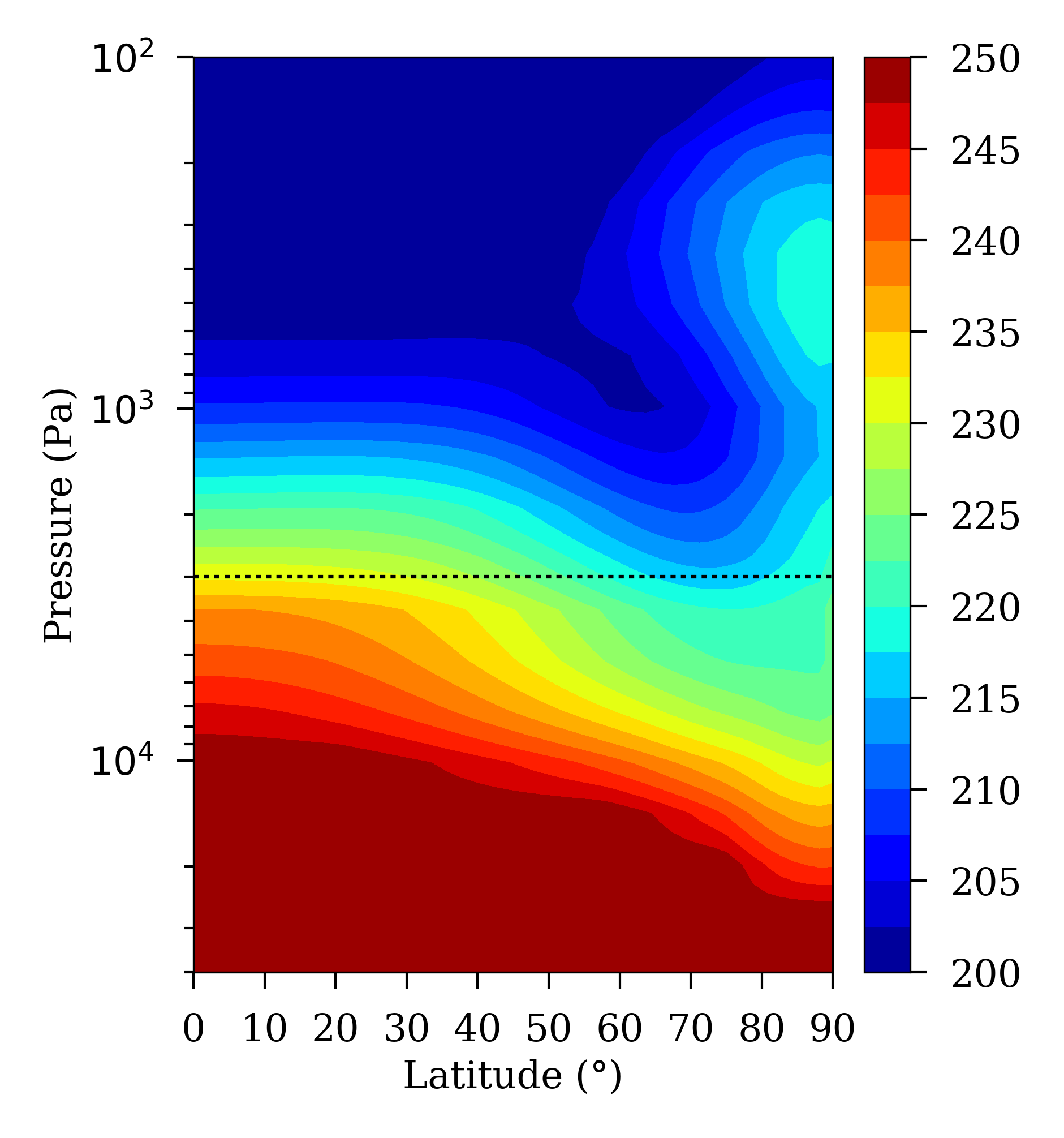}
  \end{subfigure}
  \hfill
  \begin{subfigure}[t]{0.45\textwidth}
    \centering
    \includegraphics[width=5.5cm]{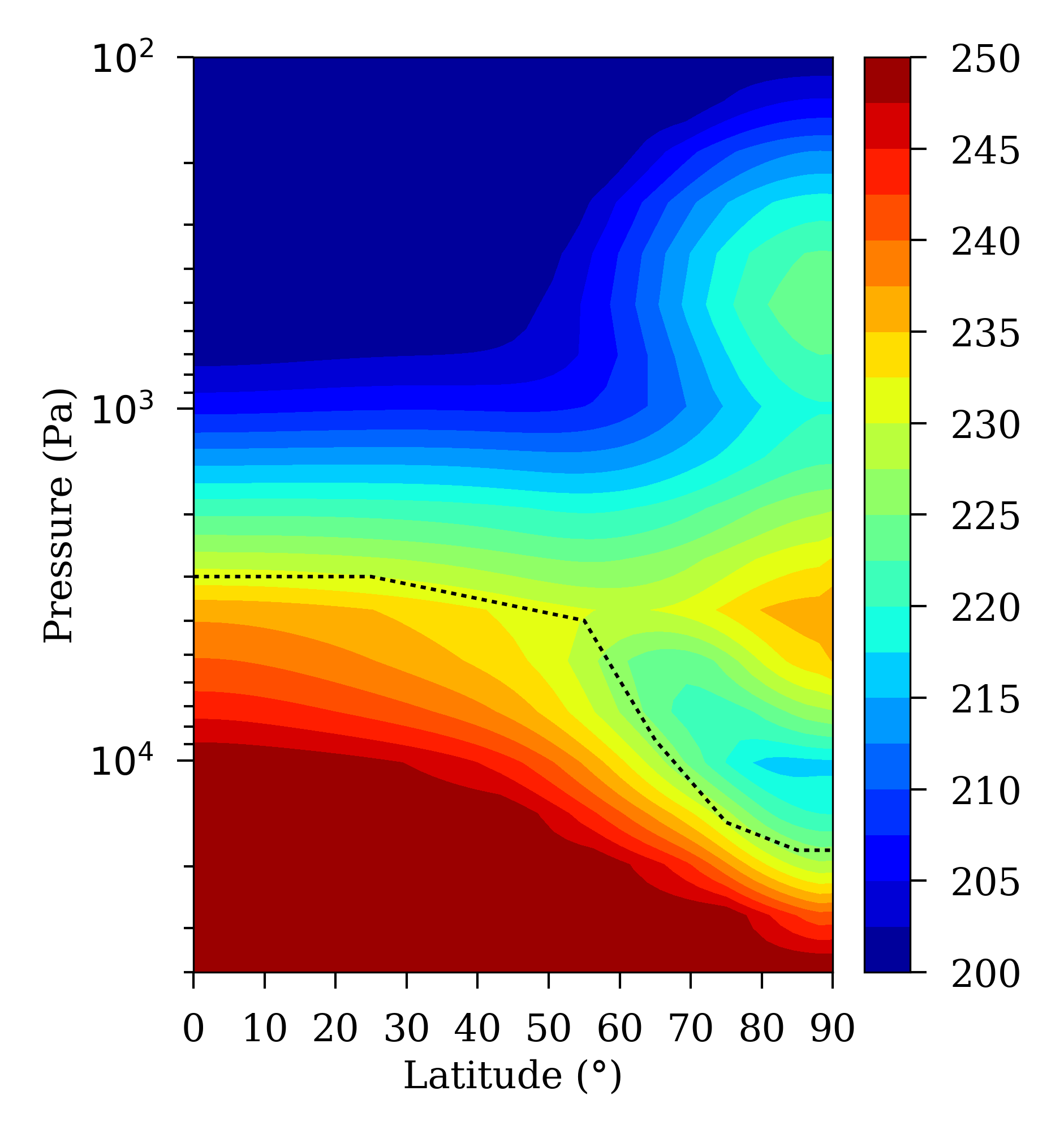}
  \end{subfigure}
\caption{Zonally and temporally averaged temperature field (K) obtained at the end of simulation (300~Vd) when considering uniform (left) and variable (right) clouds in the IR cooling computation during the last 100~Vd. Data is averaged over $360\degree$ in longitude and 2~Vd in time. Dashed line shows the cloud top altitude at 1~$\mu$m corresponding to the \cite{haus14} cloud model.}
\label{Temp1lat5lat}
\end{figure}

In the \textit{unifcld} simulation (Figure~\ref{Temp1lat5lat}a), there is a cold region around $10^{3}$~Pa and $65\degree$ that extends downward and poleward, creating a warmer region close to the pole and above that altitude level, but this cold feature is not exactly as the one observed \citep[e.g.][]{tellmann09,haus14}, it is not completely surrounded by warmer air. This structure, as well as the global thermal structure in the \textit{unifcld} simulation, is very similar to the structure obtained in previous works and presented there as the cold collar \citep{ando16, ando17, lebonnois16a}. The relation between this structure and the thermal tides in the polar atmosphere was demonstrated by \cite{ando16} and it should be similar in the \textit{unifcld} simulation presented here.

{\bf FIG~\ref{Temp1lat5lat}}

On the other hand, the cold feature seen in the average temperature map that corresponds to the \textit{varcld} simulation resembles significantly more to the observed cold collar, being more contrasted from the surrounding warmer atmosphere and located at the correct altitude and latitude. It is not completely isolated which is a characteristic different from observations. It is connected to the pole at an altitude of about $10^{4}$~Pa ($\sim$62~km) and then extends upward and equatorward. However, the connection to the pole is most likely related to the grid used currently in the model which is based on a longitude-latitude scheme with a longitudinal polar filter that reduces the latitudinal resolution and removes the high frequency variations in the polar regions. So the results at latitudes higher than $80\degree$ should be taken with caution. A new icosahedral dynamical core \citep{dubos15} will be used in the near future that should improve the robustness of the computation of the polar circulation. This should affect the fine structure of the cold collar and the inner warm core (at about $4\times10^{3}$~Pa and $80\degree$ - $90\degree$).

\begin{figure}
\centering
  \begin{subfigure}[t]{0.45\textwidth}
    \centering
    \includegraphics[width=5.5cm]{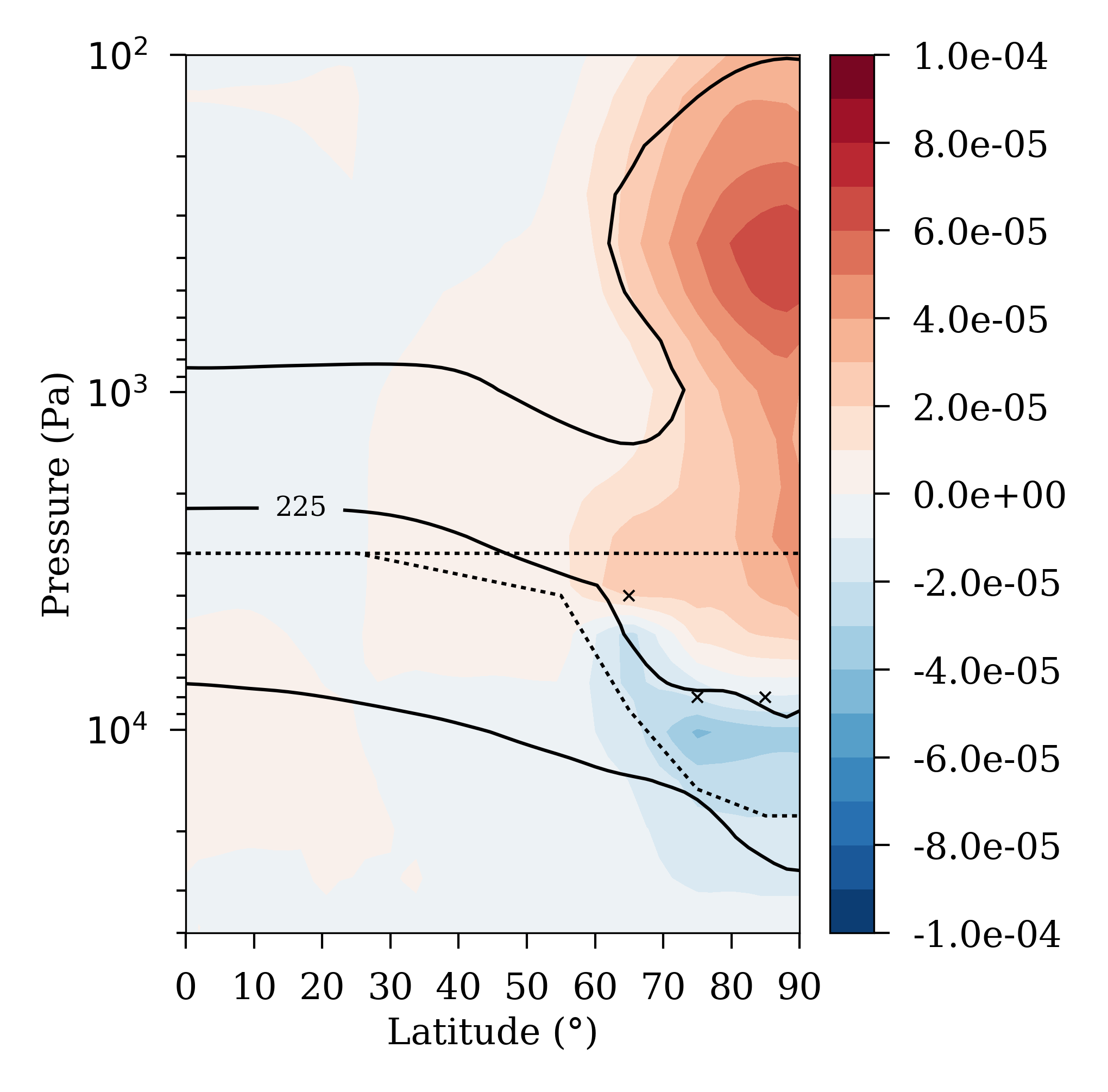}
  \end{subfigure}
  \hfill
  \begin{subfigure}[t]{0.45\textwidth}
    \centering
    \includegraphics[width=5.5cm]{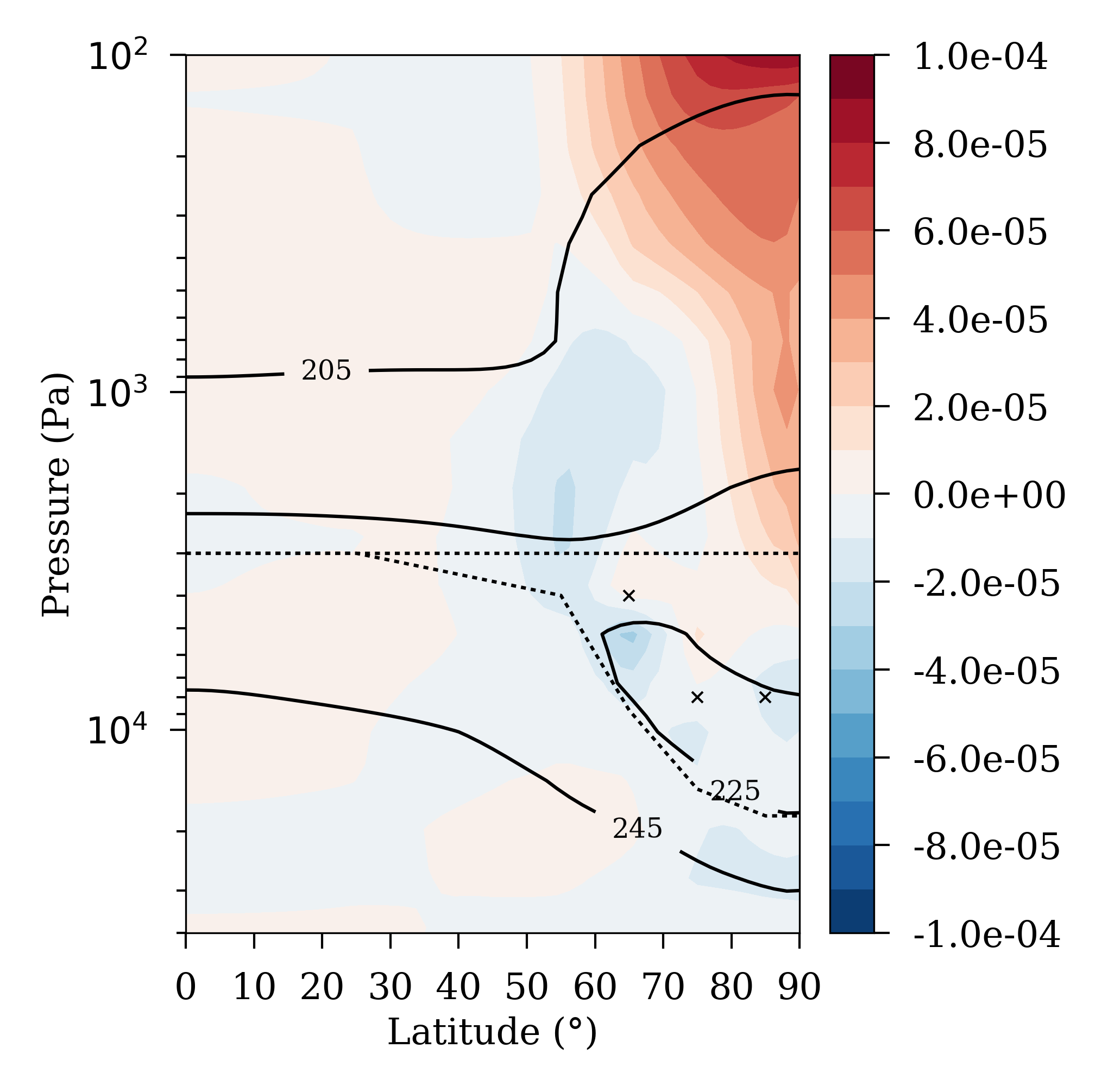}
  \end{subfigure}
\caption{Difference of the IR cooling rates (K/s) between the \textit{varcld} and \textit{unifcld} simulations as soon as the latitudinal cloud structure is implemented in the IR cooling computation (a, 0.01~Vd, instantaneous field) and 100~Vd after (b, field averaged over 2~Vd). Data is averaged over $360\degree$ in longitude. Solid lines show 205K, 225K, and 240K temperature contours of the \textit{varcld} simulation (\textit{unifcld} simulation's temperature field at 0.01~Vd is the same as the \textit{varcld} simulation and 100~Vd after is very similar to that at 0.01~Vd). Dotted lines show the cloud top at 1~$\mu$m for the uniform and variable cloud models. Crosses show the cloud upper altitude boundary (see text).}
\label{Cooling1lat5lat}
\end{figure}

To confirm the dominant influence of the cloud structure on the IR cooling and on the temperature field, the IR cooling rates of both \textit{unifcld} and \textit{varcld} simulations are compared. Figure~\ref{Cooling1lat5lat} shows the difference between \textit{varcld} and \textit{unifcld} IR cooling (IR$_{varcld}$ - IR$_{unifcld}$) at the first timestep of the simulations after the new cloud model is implemented in the IR cooling computation (0.01~Vd), so that the background temperature field has not changed and is the same for both simulations, and after 100~Vd, at the end of simulations. Note that the negative values in Figure~\ref{Cooling1lat5lat} mean IR cooling is stronger in the \textit{varcld} simulation while positive values mean it is stronger in the \textit{unifcld} simulation. The unit used in the Figure is K/s, so that $1\times10^{-4}$~K/s corresponds to $\sim$8.64~K/Earth day. This is in good agreement with \cite{leerichardson11b} who states that the IR cooling exceeds 10~K/Ed only above 100~Pa.

{\bf FIG~\ref{Cooling1lat5lat}}

From the very first moment when the cloud's latitudinal structure is taken into account in the IR radiation computation, \textit{unifcld} and \textit{varcld} simulations evolve in a completely different way with no possible convergence. The atmosphere at the \textit{varcld} simulation develops a stronger cooling region around $75\degree$ latitude and $10^{4}$~Pa altitude. 
It also develops a weaker cooling region around $85\degree$ latitude and $4\times10^{2}$~Pa altitude. During tens of Venusian days the polar atmosphere evolves in such a way that after 60~Vd it reaches a new equilibrium, although the average temperature at the cold collar oscillates slightly afterwards. Figure~\ref{Cooling1lat5lat}b shows that the stronger and weaker cooling regions in the \textit{varcld} simulation are still present 40~Vd after reaching the equilibrium, at 100~Vd. This is important because it means that the atmosphere at the newly formed cold collar will be constantly cooling by means of IR radiation in a quite stronger way than the atmosphere above it (referred to as the warm mesospheric pole hereon). As long as this new equilibrium endures, the cold collar will remain colder than its surroundings due to this stronger IR cooling.

\begin{figure}
\centering
\includegraphics[width=9.cm]{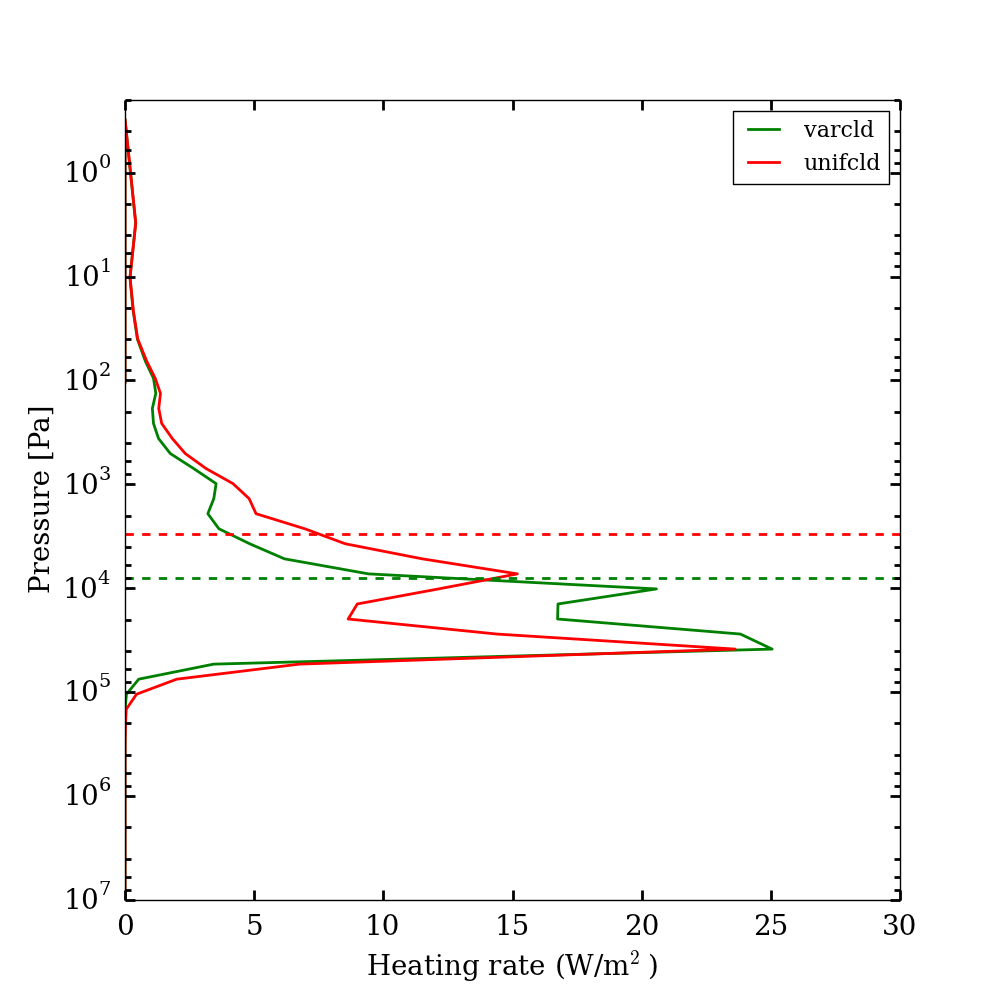}
\caption{IR energy exchanges between each layer and space at $75\degree$ latitude at the end of simulation (300~Vd, profiles averaged over $360\degree$ in longitude and 2~Vd in time) when considering uniform (red, \textit{unifcld}) and variable (green, \textit{varcld}) clouds in the IR cooling computation during the last 100~Vd. Red dashed line indicates the 1~$\mu$m cloud-top level in the \textit{unifcld} simulation, while the green dashed line indicates the cloud upper altitude boundary in the \textit{varcld} simulation.}
\label{EnergyLoss1lat5lat}
\end{figure}

Figure~\ref{EnergyLoss1lat5lat} displays the vertical profile of the IR energy exchange between different atmospheric levels and space for both \textit{unifcld} and \textit{varcld} simulations at the end of simulation (100~Vd) and at $75\degree$ latitude (cold collar region). The energy loss to space from the upper clouds ($\sim$50 - 70~km altitude) is more localized in altitude in the \textit{varcld} simulation than in the \textit{unifcld} simulation. This, as well as the decay rate of the cooling with altitude above the clouds, is inherited from the cloud model by \cite{haus14}. At high-latitude (higher than $60\degree$) the retrieval of the cloud top altitude required the estimation of another quantity, \textit{the cloud upper altitude boundary}, above which no cloud particles are considered. This cloud upper altitude boundary (located slightly above the cloud top defined at 1~$\mu$m) strongly affects the highest peak in the cooling rates, intensifying it. This peak corresponds mainly to the cooling rates in the 6-12~$\mu$m range and it is associated to the uppermost layer of cloud particles in the model. The lowest peak in the vertical profiles of Figure~\ref{EnergyLoss1lat5lat} corresponds mostly to the cooling at 20-30~$\mu$m and is associated to the base of the upper clouds, while the cooling above the clouds occurs mainly in the aisles of the 15~$\mu$m band \citep{lebonnois15}.

{\bf FIG~\ref{EnergyLoss1lat5lat}}

The position of the cold collar with respect to the cloud top (defined at 1~$\mu$m) and the cloud upper boundary is better seen in Figure~\ref{Cooling1lat5lat}. It is located just in between, coinciding with the highest peak of the cooling rates in the vertical profile. This means that the opacity at 1~$\mu$m equals unity within the upper clouds, but that there is a region with cloud particles above it where the cold collar is formed. The cold collar is, therefore, located at the uppermost layer of the clouds. Its position is, however, cloud model dependent since it is strongly linked to the cloud upper boundary defined by \cite{haus14}.

\begin{figure}
\centering
  \begin{subfigure}[t]{0.45\textwidth}
    \centering
    \includegraphics[width=5.5cm]{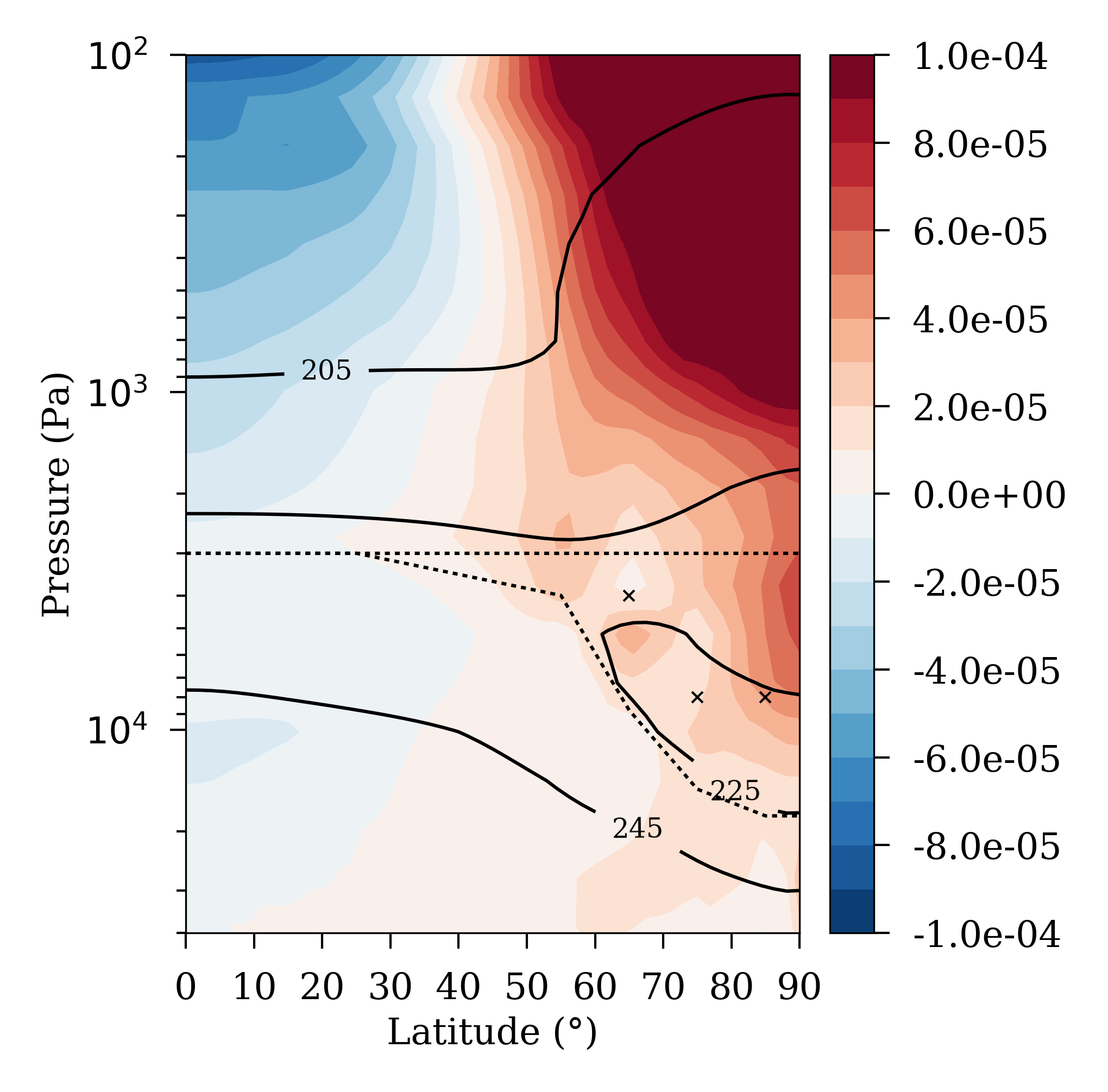}
  \end{subfigure}
  \hfill
  \begin{subfigure}[t]{0.45\textwidth}
    \centering
    \includegraphics[width=5.5cm]{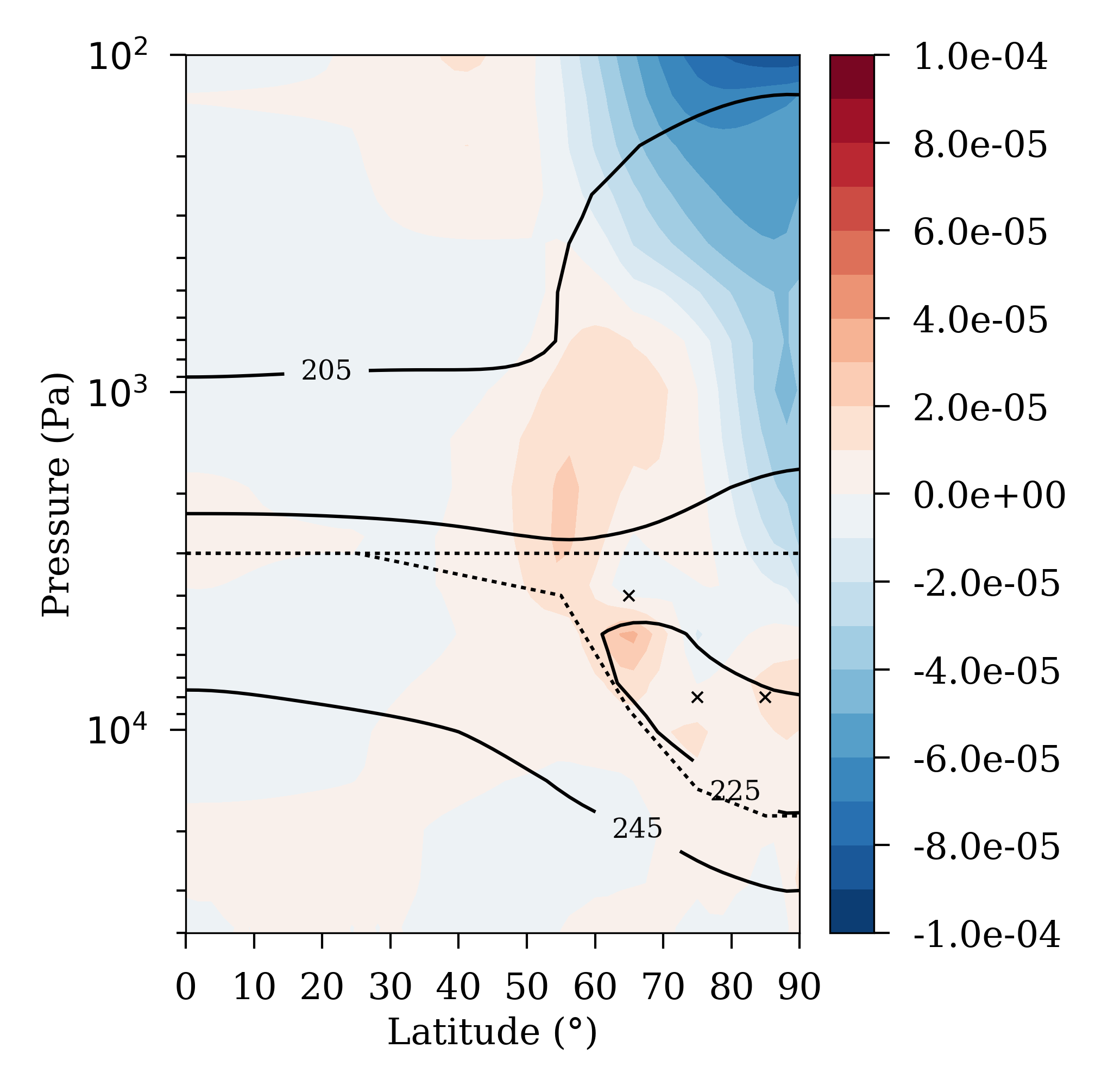}
  \end{subfigure}
  \vspace{1cm}
  \begin{subfigure}[t]{0.45\textwidth}
    \centering
    \includegraphics[width=5.5cm]{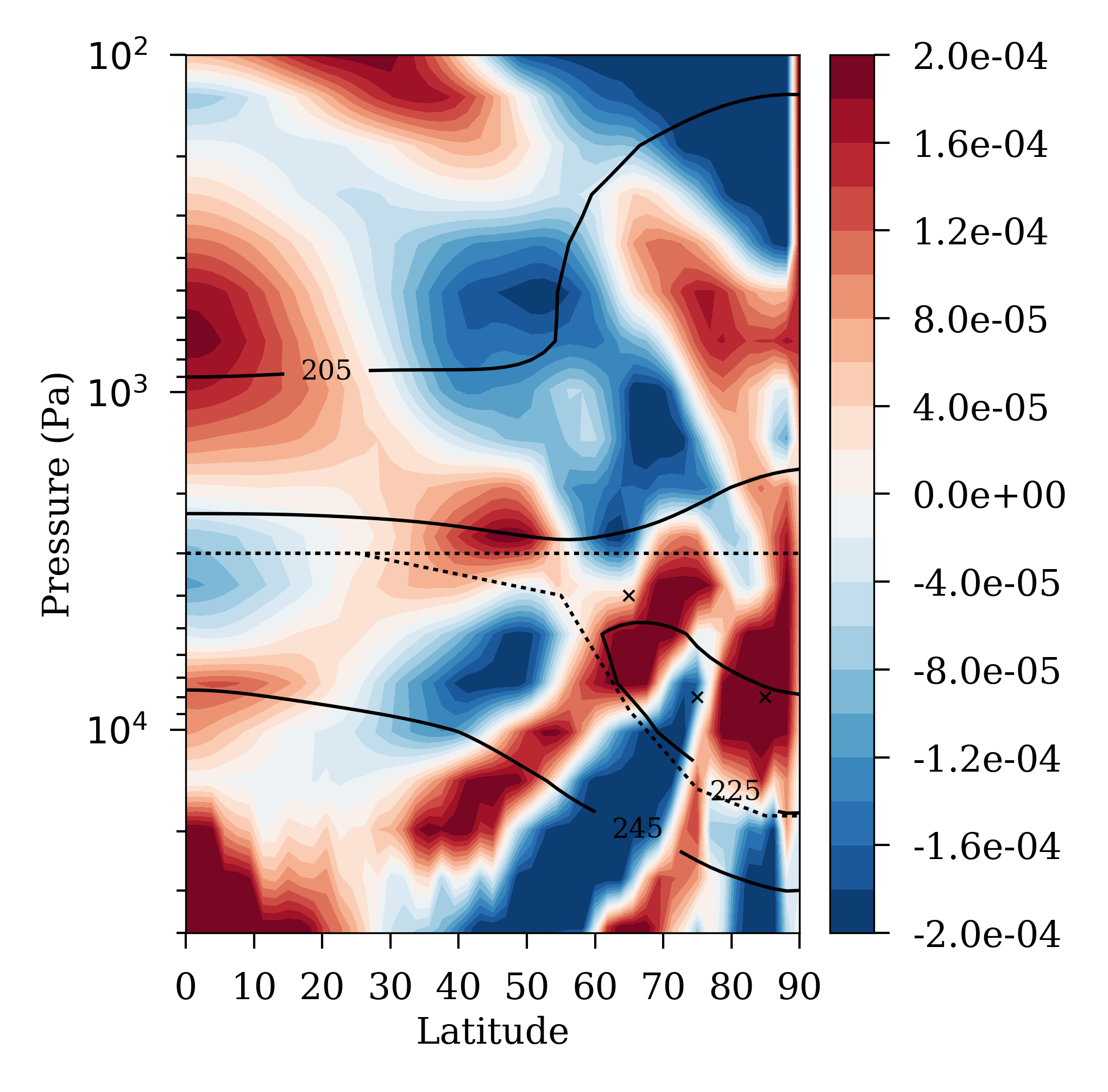}
  \end{subfigure}
  \hfill
  \begin{subfigure}[t]{0.45\textwidth}
    \centering
    \includegraphics[width=5.5cm]{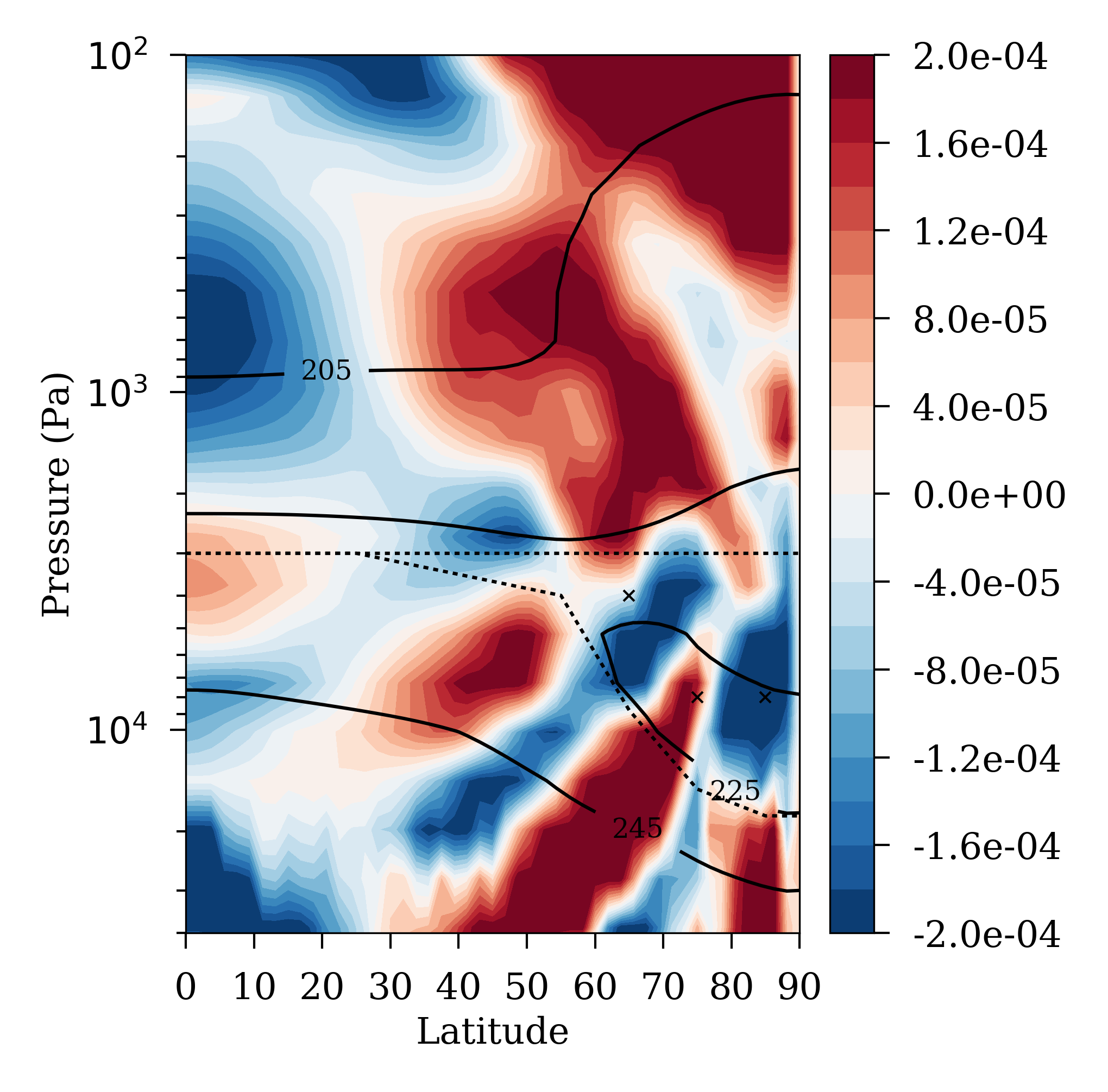}
  \end{subfigure}    
\caption{(a) dynamical heating in the \textit{varcld} simulation, (b) difference of the dynamical heating between \textit{varcld} and \textit{unifcld} simulations, (c) horizontally and (d) vertically transported dynamical heating in the \textit{varcld} simulation. All fields are plotted after 100~Vd with the latitudinal cloud structure implemented (\textit{varcld}) or not (\textit{unifcld}) in the IR cooling computation and are averaged over $360\degree$ in longitude and 2~Vd in time. Solid lines, dotted lines and crosses: same as in Fig.~\ref{Cooling1lat5lat}. Units: K/s.}
\label{DynamicsHeating}
\end{figure}

The induced modifications in the temperature structure affects the circulation and the thermal energy transport, both horizontally and vertically. A detailed analysis of the new dynamical state of the polar atmosphere and of the transport of heat and angular momentum is beyond the scope of the present paper, but the tendencies in the temperature fields due to dynamical heat transport are analyzed here to better understand the new equilibrium in which the cold collar is formed. The heating/cooling rates due to the dynamics in the \textit{varcld} simulation can be seen in Figure~\ref{DynamicsHeating}a. Dynamics is globally cooling the low-latitudes and heating the polar regions, as in the simulation with uniform cloud model and previous simulations (not shown here). Figure~\ref{DynamicsHeating}b shows the dynamical heating difference between \textit{unifcld} and \textit{varcld} simulations with positive values meaning stronger heating in the \textit{varcld} simulation. Not surprisingly, it is the inverse of Figure~\ref{Cooling1lat5lat}b since dynamical heating should compensate the radiative cooling in order to have an atmosphere in thermal equilibrium. Actually, in the cold collar area this heating due to dynamics is slightly stronger in the \textit{varcld} simulation than in the \textit{unifcld} simulation to compensate the excess of IR cooling. And in the warm pole, including the warm mesospheric pole ($85\degree$ latitude and $4\times10^{2}$~Pa) and the inner warm core ($85\degree$ latitude and $4\times10^{4}$~Pa), the dynamical heating is weaker due to the weaker IR cooling there. 

{\bf FIG~\ref{DynamicsHeating}}

In Figures~\ref{DynamicsHeating}c and \ref{DynamicsHeating}d dynamical heating in the \textit{varcld} simulation is separated between horizontally and vertically transported heating. In both panels heating is represented by red colors (positive values) and the cooling due to dynamics by blue colors (negative values). The atmosphere shows a patchy structure with dynamical heating and cooling regions interspersed with each other in both horizontal and vertical terms. These two terms are inversely related; regions where the horizontal dynamics heats the atmosphere are cooled by vertical dynamics, and vice versa. However, recall that the effects of taking into account the latitudinal variations of the clouds take place in the cold collar and warm pole areas (Figure~\ref{DynamicsHeating}b), where dynamics heats the atmosphere (Figure~\ref{DynamicsHeating}a). Above the clouds, in the warm mesospheric pole, the polar heating due to dynamics occurs by means of vertical transport (Figure~\ref{DynamicsHeating}d), but it is mitigated by horizontal transport (Figure~\ref{DynamicsHeating}c). On the other hand, in the cold collar and inner warm core areas, horizontal transport mostly heats the area (although a region of dynamical cooling is appreciated between the main two heating regions). This heating is mitigated by vertical heating transport.

\begin{figure}
\centering
  \begin{subfigure}[t]{0.45\textwidth}
    \centering
    \includegraphics[width=5.5cm]{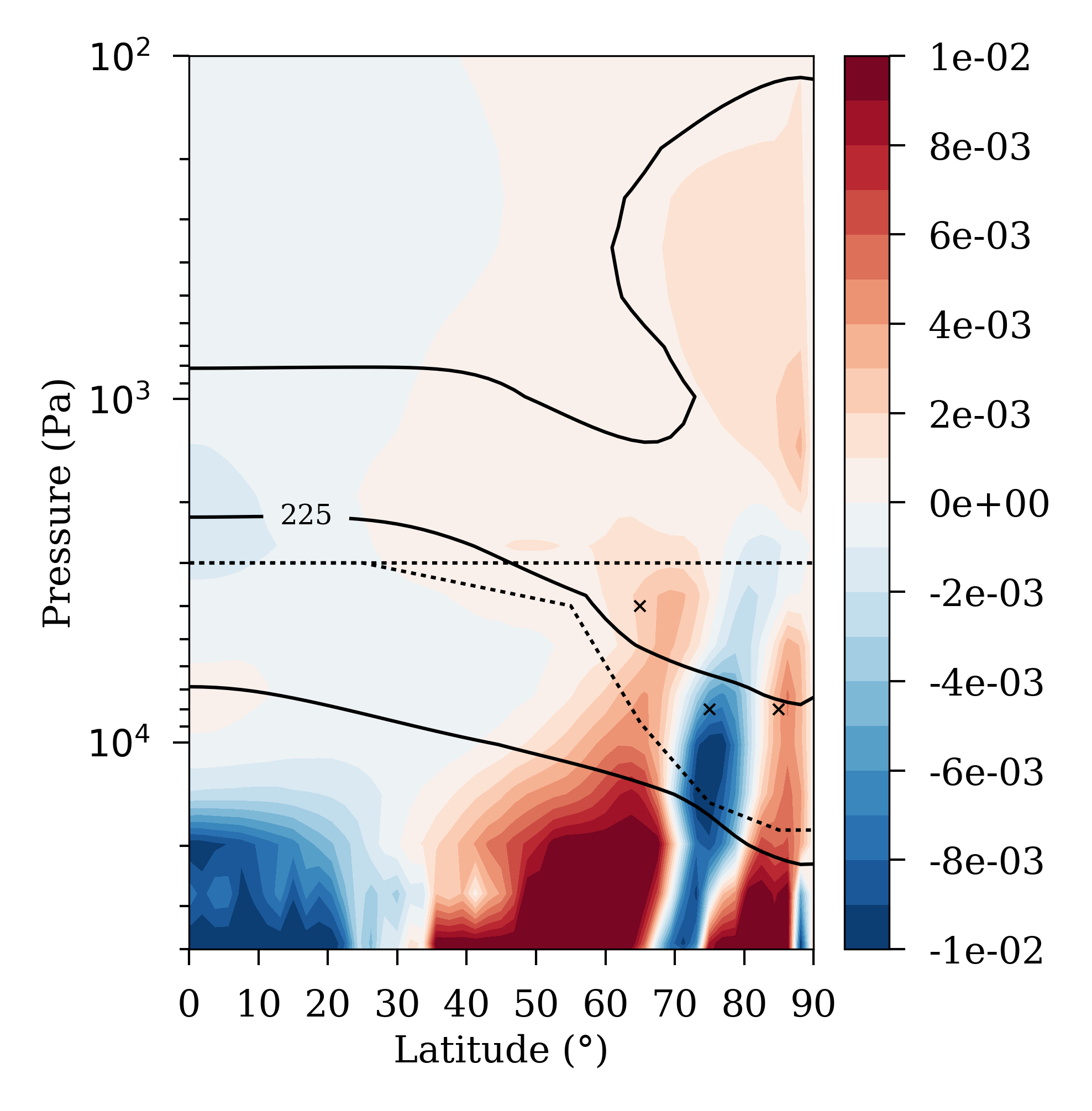}
  \end{subfigure}
  \hfill
  \begin{subfigure}[t]{0.45\textwidth}
    \centering
    \includegraphics[width=5.5cm]{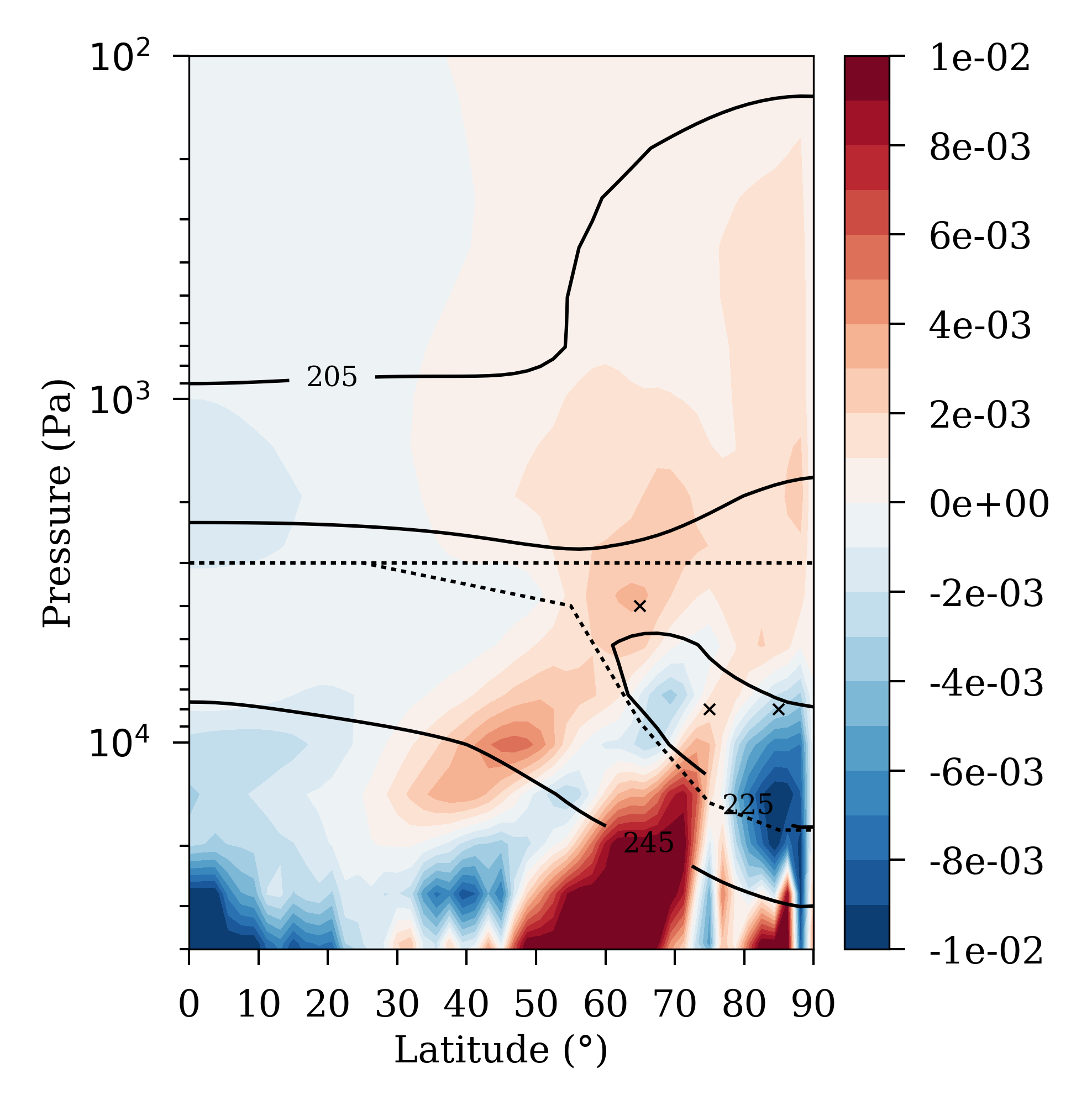}
  \end{subfigure}
\caption{Zonally and temporally averaged vertical wind velocity (Pa/s) at the end of simulation (300~Vd), when considering uniform (a) and variable (b) clouds in the IR cooling computation during the last 100~Vd. Data is averaged over $360\degree$ in longitude and 2~Vd in time. Solid lines, dotted lines and crosses: same as in Fig.~\ref{Cooling1lat5lat}.}
\label{VertWind1lat5lat}
\end{figure}

Studying the variation of the circulation, particularly of the zonal and meridional winds, is not trivial due to the high variability present in the polar regions. This study of the horizontal distribution of the dynamical polar atmosphere in the \textit{varcld} simulation will be presented in a subsequent paper together with the new wave activity. However, the variation of vertical wind is interesting and noteworthy. Figure~\ref{VertWind1lat5lat} shows the vertical wind velocity fields in both \textit{unifcld} and \textit{varcld} simulations (in Pa/s units, so negative values mean upwelling motion). Although the main differences take place at the clouds and levels below, an altered subsidence structure that extends down to $50\degree$ latitude is seen in the \textit{varcld} simulation, responsible of the vertical heating seen in Figure~\ref{DynamicsHeating}d. Ascending air is found in the cold collar between $60\degree$ and $70\degree$ latitudes and at about $10^{4}$~Pa altitude for the \textit{varcld} simulation which is associated with the dynamical cooling seen in Figure~\ref{DynamicsHeating}d. A reversal of the vertical circulation in the $70\degree$ - $90\degree$ polar region from \textit{unifcld} to \textit{varcld} simulations is also clearly seen.

{\bf FIG~\ref{VertWind1lat5lat}}

Hence, according to the results of the new IPSL Venus GCM, the cold collar feature observed in Venus' polar atmosphere is formed by the infrared cooling region generated by the decrease of the cloud top altitude poleward and the latitudinal variation of the particles modes. Since the radiative forcing cools the area while the dynamics heats it, resulting in a region characteristically cold, it seems that the driver of the cold collar is the radiative transfer rather than dynamics. However, the presence of the cold collar strongly influences the dynamics, thus affecting the polar circulation of Venus. Even though the cloud distribution, radiative transfer, and dynamics form a coupled system in reality and, therefore, any change in one of the mentioned aspects will bring along a change in the other aspects, this coupling is not yet present in our GCM and we can not characterize it in detail.

The temporal evolution observed in the modeled polar temperature maps at $7\times10^{3}$~Pa altitude (not shown here) presents a cold collar that is displaced from the pole and rotates around it. The obtained rotation period of 5.85 Earth days agrees with the period found by \cite{luz11} for the drift of the rotation center of the vortex around the pole. This result, together with the structure of the cold feature in the average temperature field, indicates that the current IPSL Venus GCM models satisfactorily the subpolar latitudes. However, the current polar latitudes' treatment is not ideally suited and, therefore, the modeled inner core does not show neither a clear rotation around the pole nor detailed horizontal structure within it, as the observed polar vortex does.


\section{Conclusions}\label{Conclusions}

The previous IPSL Venus GCM \citep{lebonnois16a} has been modified in order to improve the radiative transfer, update the cloud model, and take into account the latitudinal variation of the clouds (by means of cloud top altitude and different modes' abundances) in the solar heating profiles and infrared cooling computation. The new cloud model used in the current version is described by \cite{haus14} and shows that the cloud top altitude decreases towards the pole in Venus's atmosphere and that the different particle modes' abundances change from equator to pole.

The zonally and temporally averaged zonal wind and temperature fields have been improved showing now a better agreement with the observations. Particularly, the equatorial jet's intensity has decreased compared to the mid-latitude jets' intensity, the zonal wind velocities between 40 and 60~km altitude are faster than previously, and the averaged temperature values below (resp. above) the clouds are now warmer (resp. colder), thus closer to the VIRA reference profile.

The main result, however, is the appearance of a feature in the averaged altitude-latitude temperature map that resembles the subpolar cold collar observed in Venus's atmosphere. The atmosphere between $60\degree$ and $85\degree$ shows a thermal minimum at $2\times10^{4} - 4\times10^{3}$~Pa ($\sim62 - 66$~km) altitude, being $\sim$20~K colder than the equator and about 10~K colder than the pole. The coldest region shows values about 220~K, in good agreement with observed values, but the inner warm core shows an average value of 230~K that is colder than the observed values. The structure simulated within the inner core is smoother than that obtained by \cite{yamamototakahashi15} and \cite{ando16, ando17}, but the altitude level where the cold collar is reproduced and the modeled temperature values are more consistent with observations.

According to our analysis the cloud's latitudinal structure is an essential element in the formation of the cold collar via the IR cooling rates. A strong cooling region at $\sim75\degree$ and $10^{4}$~Pa is developed as soon as the new cloud structure is implemented in the IR computations and it remains cooling strongly after a new equilibrium is reached and the cold collar is formed. A region of weak IR cooling is also developed above the polar clouds just where the warm mesospheric pole is observed.

The position of the cold collar is cloud model dependent since it is strongly linked to the cloud upper altitude boundary (level above which no particles are considered) defined by \cite{haus14} at high latitudes. The IR cooling peak is located right below this cloud upper boundary, above the cloud top altitude. This means that the cold collar is located at the uppermost layer of the clouds, in a region with cloud particles but above the level where the opacity equals one at 1~$\mu$m.

Once the atmosphere reaches a new thermal equilibrium, the radiative cooling is compensated by heating due to dynamics. Consequently, the dynamical heating is stronger at the cold collar area than in the warm mesospheric pole ($85\degree$ and $4\times10^{2}$~Pa). Our analysis points to vertical transport (associated to a subsidence motion that extends down to $50\degree$) as the main source of dynamical heating in the warm mesospheric pole and to horizontal transport at the cold collar.

Since the radiative forcing cools the area while the dynamics heats it, resulting in a region characteristically cold, it seems that the driver of the cold collar is the radiative transfer rather than dynamics.


\section*{Acknowledgment}

{\small This work was supported by the Centre National d'Etudes Spatiales (CNES). IGL was supported by CNES postdoc grant. GCM simulations were done at CINES, France, under the project n$^\circ$11167. We also thank E. Millour for his technical support with the model.}


\renewcommand{\baselinestretch}{1}

\clearpage
\bibliography{./GL18_paper_arxiv}
\bibliographystyle{apalike}





\clearpage
\begin{table}[ht]
\caption{Chosen inputs for the simulations.}
\label{paramsimus}
\vspace{5pt}

\begin{tabular}{l|cc|cc|ccc|cc}
\hline
\hline
& \multicolumn{2}{c|}{Solar heating} & 
  \multicolumn{2}{c|}{Correlated-k set} &
  \multicolumn{3}{c|}{Cloud model (IR)} &
  \multicolumn{2}{c}{Continuum (3-7$\mu m$)} \\
& C86$^a$ & H15$^b$ & L16$^c$ & updated & Z07$^d$ & H13$^e$ & H14$^f$ & 
  L16$^c$ & updated$^g$ \\
\hline
Initial state & $\times$ & & $\times$ & & $\times$ & & & $\times$ & \\
0-200 Vd      & & $\times$ & $\times$ & & & $\times$ & & & $\times$ \\
200-300 Vd: & & & & & & & & & \\
- \textit{unifcld} & & $\times$ & & $\times$ & & $\times$ & & & $\times$ \\
- \textit{varcld} & & $\times$ & & $\times$ & & & $\times$ & & $\times$ \\
\hline
\hline
\end{tabular}
$^a$ \citet{crisp86} \\
$^b$ \citet{haus15}, with additional tuning in lower haze area (see text) \\
$^c$ \citet{lebonnois16a} \\
$^d$ \citet{zasova07} \\
$^e$ \citet{haus13}, clouds taken as uniform with latitude \\
$^f$ \citet{haus14}, clouds taken as variable with latitude \\
$^g$ see text
\end{table}



\clearpage

\clearpage

\clearpage

\clearpage

\clearpage

\clearpage

\clearpage

\clearpage

\clearpage

\clearpage


\end{document}